\documentclass[a4paper,11pt]{article}
\usepackage{jheppub} 

\usepackage{amsmath,amsfonts,bbm,mathrsfs,amssymb}

\usepackage{color}
\usepackage[table,xcdraw,dvipsnames]{xcolor}
\usepackage{bm} 
\usepackage{hyperref}
\hypersetup{colorlinks=true,urlcolor=blue,anchorcolor=blue,citecolor=blue,filecolor=blue,
            linkcolor=blue,menucolor=blue, linktocpage=true,pdfproducer=medialab}
\usepackage[english]{babel}

\hypersetup{
    pdftoolbar=true,        
    pdfmenubar=true,        
    pdffitwindow=false,     
    pdfstartview={FitH},    
    pdftitle={Unveiling atomic electron motion effects in e+e− high 
    energy collisions at NA64},
    pdfsubject={High energy physics}, 
    pdfauthor={F. Arias Aragon, G. Grilli di Cortona, E. Nardi},
    pdfkeywords={di-muon production,} {atomic electron velocities},
    pdfnewwindow=true,
    colorlinks=true,
    linkcolor=blue,
    citecolor=blue,
    filecolor=blue,
    urlcolor=blue,
    linktocpage=true
}

\def\beq{\begin{equation}}  
\def\eeq{\end{equation}}
\def\beqa{\begin{eqnarray}}  
\def\eeqa{\end{eqnarray}}

\newcommand{\ba} {\begin{equation}\begin{aligned}}
\newcommand{\ea} {\end{aligned}\end{equation}}

\newcommand{\bg} {\begin{equation}\begin{gathered}}
\newcommand{\eg} {\end{gathered}\end{equation}}

\newcommand{\cM}{\mathcal{M}}

\def\({\left(}
\def\){\right)}
\def\[{\left[}
\def\]{\right]}

\def\eg{\hbox{\it e.g.}{}}

\newcommand{\GeV}{\,\mathrm{GeV}}

\newcommand*{\INFNFR}{Istituto Nazionale di Fisica Nucleare, Laboratori Nazionali di Frascati, C.P. 13, 00044 Frascati, Italy}
\newcommand*{\LNGS}{Istituto Nazionale di Fisica Nucleare, Laboratori Nazionali del Gran Sasso, Assergi, 67100, Italy}
\newcommand*{\NICPB}{Laboratory of High Energy and Computational Physics, NICPB, R\"avala 10, 10143, Tallin, Estonia}

\title{Unveiling atomic electron motion effects in $e^+e^-$ high energy collisions at NA64}

\author[a]{Fernando Arias-Aragón,}
\author[b]{Giovanni Grilli di Cortona,}
\author[a,c]{Enrico Nardi,}
\affiliation[a]{\INFNFR}
\affiliation[b]{\LNGS}
\affiliation[c]{\NICPB}

\emailAdd{fernando.ariasaragon@lnf.infn.it, giovanni.grilli@lngs.infn.it, enrico.nardi@lnf.infn.it}

\abstract{Atomic electron motion is responsible for well-studied effects observed in low-energy (MeV-scale) processes. Recently, interest in this phenomenon has also emerged within the 
high-energy physics community, due to its potential to  increase significantly the center-of-mass energy in fixed-target experiments. However, direct experimental evidence of this effect in high energy collisions has yet to be observed.
We argue that a striking manifestation of atomic electron momenta could be revealed
by the NA64 experiment at CERN during the proposed run 
with a 40\,GeV positron beam. 
At this energy, $\mu^+\mu^-$ production  via positron annihilation 
on  electrons at rest is kinematically forbidden. The detection of $\mu^+\mu^-$ pairs from the annihilation channel  would thus constitute direct evidence of an increase in the center-of-mass energy resulting from atomic electron motion. 
We also investigate the expected signatures for the proposed 60\,GeV 
run, as well as for the data already collected at 70\,GeV. 
Intriguingly, in both these cases, the predicted number of $\mu^+\mu^-$ pairs from positron annihilation is reduced compared to the electron-at-rest approximation.}

\begin{document}
\maketitle
\flushbottom

\section{Introduction}
The determination of the atomic and electronic structure of materials is of fundamental importance in solid-state physics and chemistry. A particularly important quantity in this context is the atomic electron momentum distribution, which carries critical information since it is intimately related to the square of the electron wave functions in momentum space. These wave functions,  in turn,
are the Fourier transforms of the corresponding position-space wavefunctions. 
Several well-established and widely used experimental techniques probe this distribution by measuring the Doppler broadening of photon lines induced by electron motion.
This effect is observed, for example, in Compton scattering experiments (see, e.g., Ref.~\cite{Williams:01011983}) and in positron annihilation spectroscopy, which involves precise measurement of the  511\,keV photon line resulting from the annihilation of stopped positrons  (see, e.g., Ref.~\cite{Ghosh:2000}).

A class of high-energy physics experiments involves firing high-energy 
positron beams at a fixed target, followed by the detection of particles 
produced in positron annihilation off atomic electrons~\cite{Battaglieri:2021rwp}. 
A key advantage of this approach is the significantly higher particle density in fixed targets, which allows for interaction rates many orders of magnitude greater than that achievable in head-on beam–beam collisions.
For this reason  such experiments are particularly well suited for probing feebly interacting particles (FIPs)~\cite{Antel:2023hkf}, which are characterized by sub-electroweak-scale masses and
highly suppressed production cross sections.\footnote{Similar considerations apply also to electron~\cite{Marsicano:2018krp} and proton beam dump experiments~\cite{Celentano:2020vtu}. In these cases  positrons are created as secondary particles in electromagnetic showers.}
It was recently recognized that a proper interpretation of results from this class of experiments 
must account for the fact that atomic electrons are not at rest, but instead exhibit a distribution of velocities~\cite{NewDirections, Nardi:2018cxi}. Ultimately, this is a direct consequence of Heisenberg’s uncertainty principle, which dictates that localized electrons must have a non-trivial momentum distribution.
In particular,  for the highly localized inner-shell electrons of high-$Z$ elements, the momentum distribution  remains significant up to relativistic values, implying that high-$Z$ atoms can effectively act as electron accelerators, 
 enhancing the center-of-mass (c.m.) energy of the collision. On the one hand,   this effectively extends  the mass reach  of FIPs searches  by a factor   of a few
 compared to the free electron-at-rest (FEAR) approximation~\cite{Arias-Aragon:2024qji,Arias-Aragon:2025kiz}.  On the  other hand, this phenomenon has been proposed as a novel pathway  for measuring the hadron cross section in  $e^+e^-$ annihilation up to $\sqrt{s} \sim 1\,$GeV, using a positron beam with energies as low as 12\,GeV, such as that foreseen at 
 Jefferson Laboratory  (JLab)~\cite{Arias-Aragon:2024gpm}.

%

While it is certainly worthwhile to further develop these proposals and explore other processes where this phenomenon may be relevant, it is important to emphasize that direct experimental evidence of bound-electron momentum effects in high-energy collisions has yet to be observed.\footnote{A possible exception is the $14\%$ discrepancy in polarization measurements of the $\sim 46\,$GeV SLAC electron beam performed with Compton and M{\o}ller polarimeters~\cite{Swartz:1994yv}, 
which has been ascribed to the intrinsic momenta of the target electrons, a phenomenon 
known as the Levchuk effect~\cite{Levchuk:1992np}.
}
In this paper, we argue that the NA64 experiment at CERN, 
thanks to his high-$Z$ (Pb/Sc) active target, while pursuing its dedicated search for light dark matter with positron beams~\cite{NA64:2023ehh,Bisio:2887649,NA64:2025rib},  also offers an ideal setting for collecting direct and striking evidence of bound-electron momentum effects.
In particular, during the planned run with a 40\,GeV positron beam~\cite{Bisio:2887649}, muon pairs cannot be produced via $e^+$ annihilation on stationary electrons, as the beam energy lies below 
the di-muon production threshold:
\begin{equation}
E_\mathrm{th}=\frac{2 m_\mu^2}{m_e}-m_e\simeq 43.7 \,\mathrm{GeV},
\label{eq:threshold}
\end{equation}
where $m_\mu$ and $m_e$ are the muon and electron mass, respectively.
Any di-muon signal from $e^+e^-$ annihilation observed in the 40\,GeV NA64 run must therefore originate from head-on collisions of the positrons with atomic electrons carrying sufficiently large momentum. 
Moreover, as we will argue, atomic electron momentum effects can also manifest 
as deviations in the measured number of di-muon events relative to the FEAR approximation 
in runs where the beam energy is above $E_\mathrm{th}$, such as the planned  60\,GeV~\cite{Bisio:2887649} and 70\,GeV\cite{NA64:2025rib} NA64 runs. 

In Section~\ref{sec:overview},
we recall key aspects of the NA64 experimental setup and suggest strategies for distinguishing di-muons 
events arising from $e^+e^-$ annihilation from those produced in hard positron-nucleus interactions. 
In Section~\ref{sec:production}, we show how to incorporate the atomic electrons momentum distribution into the calculation of the rate for di-muon production  in $e^+e^-$ annihilation. Section~\ref{sec:uncertainties}, 
discusses the  theoretical uncertainties associated with the evaluation of the electron momentum distribution. The significance  
of our findings is discussed in Section~\ref{sec:results}. Finally, in  Section~\ref{sec:conclusions} we draw our conclusions.

\section{Overview of the experimental setup}
\label{sec:overview}
 The measurement of atomic electron momentum effects 
 in high energy collisions that we are suggesting can leverage the positron beam from the CERN 
 H4 beamline impinging on the NA64   active target. 
The active target is a 40 radiation lengths ($X_0$) high-efficiency
electromagnetic calorimeter (ECAL) made by a matrix of 
modules  assembled as a  sandwich of 
 inhomogeneous lead and plastic scintillator (Pb/Sc)  plates. 
The first $4X_0$ form a pre-shower section (ECAL0), while the remaining $36X_0$ constitute the main sampling section (ECAL1). 
Unfortunately, the current configuration cannot perform a precise $\mu^+\mu^-$ invariant mass measurement. Although a tracking system is present downstream of the ECAL and hadronic calorimeter (HCAL), muons must traverse both detectors, incurring in significant energy loss by ionization and multiple scattering. These effects degrade momentum resolution and alter their trajectories, preventing an accurate reconstruction of the dimuon invariant mass. The NA64 experiment can, however, count the total number of di-muon events in ECAL0, where virtually all di-muon events from positron annihilation are produced. This is because, after traversing four radiation lengths, the positrons lose enough energy that they no longer contribute significantly to this type of event.

Besides the $e^+e^- \to \mu^+\mu^-$ annihilation channel, di-muon production can also occur through other processes, such as the conversion of bremsstrahlung photons into $\mu^+\mu^-$ pairs in the nuclear electric field, or via the trident process, that is the pair production of $\mu^+\mu^-$ from a virtual photon emitted in electron–nucleus scattering.
These alternative production mechanisms can be distinguished from the annihilation channel by two specific features. First, the primary positron survives the interaction and, unlike muons, will deposit most of its energy within the first few radiation lengths of the detector. Consequently, vetoing events with substantial energy deposition in ECAL0 can provide an effective method for selecting di-muon events originating from positron annihilation. Second, di-muon production via the alternative processes is identical for both electron and positron beams. Therefore,  
 backgrounds to di-muon production via annihilation can be effectively subtracted by measuring the  di-muon event rate in electron mode.

\section{Di-muon production}  
\label{sec:production}
\begin{figure}[t!]
    \centering
        \includegraphics[width=0.75 \textwidth]{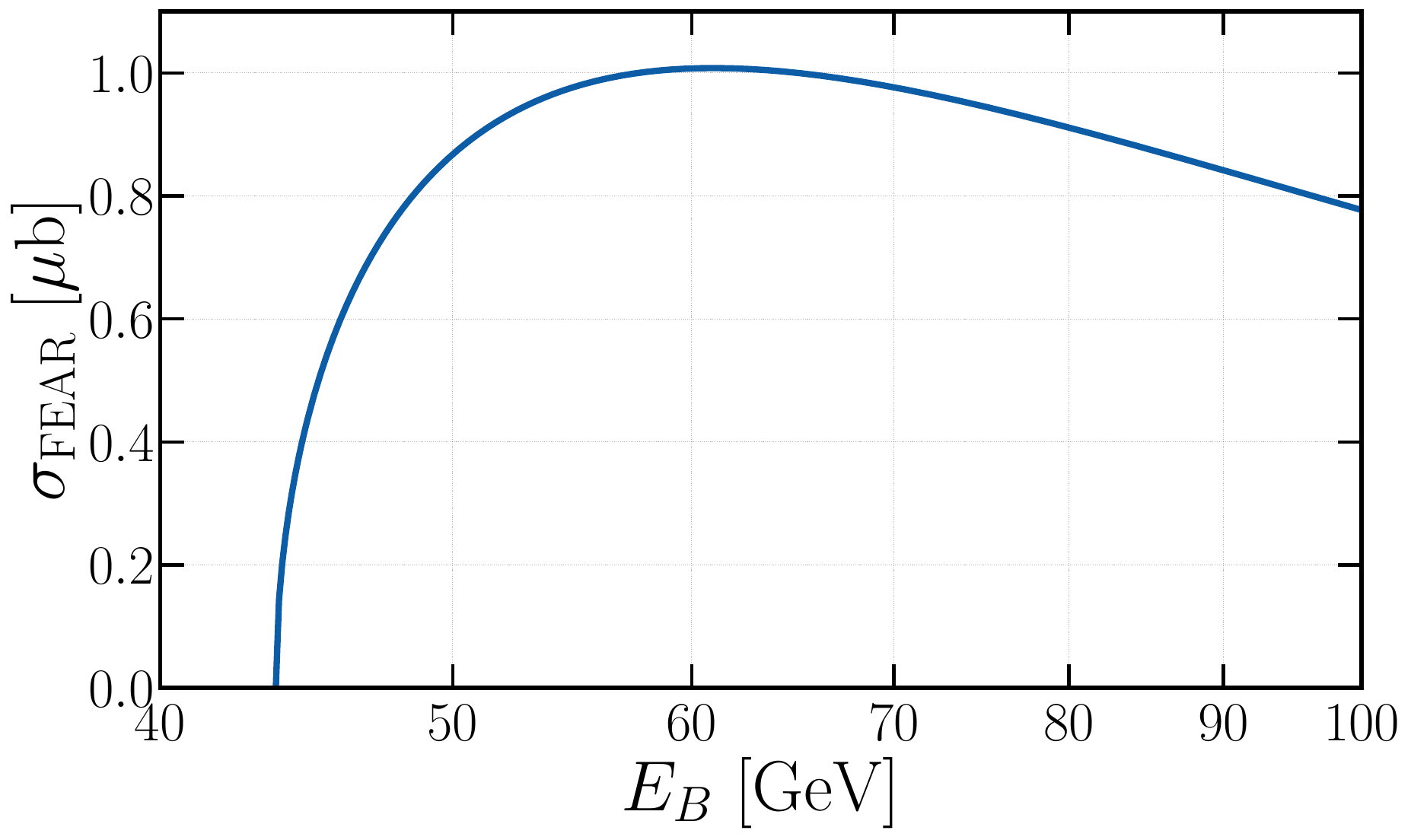}
    \caption{Cross section for the $e^+e^- \to \mu^+\mu^-$ process in the FEAR approximation as a function of the beam energy $E_B$.}
    \label{fig:FEARxs}
\end{figure}

%
The total cross section for the $e^+e^- \to \mu^+\mu^-$  process in the 
FEAR  approximation~\cite{ParticleDataGroup:2020ssz} is shown in Fig.~\ref{fig:FEARxs}, and is  
given by
\begin{equation}
\sigma=\frac{4\pi\alpha^2}{3\,s}\left(1+\frac{2\,m_\mu^2}{s} \right)\sqrt{1-\frac{2\,m_\mu^2}{s}},
\end{equation}
where $s=2m_e(E_B+m_e)$ is the centre of mass energy squared, with $E_B$ the total positron energy in the laboratory frame.
However, the electrons in the target are neither free nor at rest; they are spatially confined within atomic orbitals, which implies a non-trivial probability distribution for their momenta. The physical process involves the annihilation of a positron with four momentum $p_B=(E_B, 0, 0, p_B)$ on an atomic electron belonging to an orbital labeled by quantum numbers $(n,\,\ell)$, with momentum-space wavefunction $\phi_{n \ell}(\mathbf{k}_A)$. The final state is $\mu^-(p_1)+\mu^+(p_2)$. 
By neglecting the electron binding energy and  assuming an isotropic electron momentum distribution:
\ba
E_A \simeq  & \ m_e\,, \\
n(k_A)=& \ \sum\nolimits_{n\ell}\left|\phi_{n\ell}(k_A)\right|^2\,,
\ea
%
the differential cross section can be written as (see Refs.~\cite{Arias-Aragon:2024qji,Arias-Aragon:2024gpm} for details):
\ba
\label{eq:d2sigma}
\frac{d^2 \sigma}{d s \,dc_{\theta_2}}  = 
 \int_{m_\mu}^\infty \!\!  dE_2  
\int_0^\infty \!\! dk_A \; \frac{\left|\cM\right|^2\,n(k_A)} {2^9 \pi^5}\frac{  (2\pi-\arccos{d})\, \Pi\left(\frac{d}{2}\right)}
{p_B |E_Bk_A c_{\theta_A}-E_{k_A}p_B|\, s_{\theta_2}s_{\theta_A}\, \sqrt{1 - d^2}} \,,
\ea
where $E_2$ is the energy of the final-state muon, 
 we have used spherical coordinates: $d^3 k_A = k_A^2 d k_A d c_{\theta_{\!A}}  d \varphi_A $
and $d^3 p_2 = p_2^2 d p_2\,d c_{\theta_2} d \varphi_2$, and 
we have defined
\beq\nonumber 
d=(a-\mathcal{E}^2)/b,
\eeq
with 
\begin{eqnarray}
\nonumber
a \! &=&\! p_2^2+m_\mu^2+k_A^2+p_B^2+2k_A p_B c_{\theta_A}\!-2 p_2 c_{\theta_2} (p_B+k_A  c_{\theta_{\!A}}),\\
\nonumber
b \! &=& \! 2k_A  p_2 s_{\theta_2} s_{\theta_{\!A}},  \\
\nonumber
\mathcal{E} \! &=& \! E_A+E_B-E_2\,.
\end{eqnarray}
%
The rect function $\Pi(d/2)$ enforces the kinematic condition $|d|<1$ which 
ensures energy conservation, and also defines a minimum electron momentum $k_A^\mathrm{min}$.  
The energy $E_{k_A}=\sqrt{k_A^2+m_e^2}$ arises from the normalization of the one-particle free-electron states. 
The differential cross section depends on the spin-averaged matrix element  between free particle states
\ba
|\cM|^2=&\frac{32\pi^2\alpha^2}{(m_\mu^2+ P_1\cdot P_2)^2}\biggl\{m_e^2 \bigl[2m_\mu^2+(P_1\cdot P_2)\bigr] + m_\mu^2(K_A\cdot P_B) \\ 
+& (K_A\cdot P_1)(P_B\cdot P_2)  
+ (K_A\cdot P_2)(P_B\cdot P_1) \biggr\},
\ea
where capital letters denote free particle four-momenta (e.g. 
$K_A=(E_{k_A},\bm{k}_A$)), constrained by the three-momentum conservation condition~\cite{Essig:2011nj}.
The electron momentum distribution $n(k)$ can be reconstructed from 
experimental measurements as well as from theoretical computations 
of the isotropic Compton profile (CP)~\cite{Williams:01011983}
$J(k)$ via the relation $n(k)=-(2\pi)^2\,k^{-1} \,dJ(k)/dk$. 
Specifically, we obtained $n(k_A)$ for lead up to 
electron momenta $k_A\sim 78$~a.u.$\;\sim 2.9\times 10^{-4}$\,GeV from the CPs given in Refs.~\cite{MENDELSOHN1974521,BIGGS1975201,Pattison_1979,Timms1993}. For larger momenta, 
up to $k_A\sim 2700$~a.u.$\;\sim 10^{-2}$\,GeV, we  computed $n(k_A)$ including  
 contributions from core orbitals up to the $4d$ shell, using 
the  DBSR-HF code~\cite{ZATSARINNY2016287}.

The number of $e^+e^-\to \mu^+\mu^-$ events produced in a thick target of length $T$ in units of radiation lengths is given by~\cite{Nardi:2018cxi}
\ba
    \frac{N_{\mu\mu}}{N_{e^+\mathrm{oT}}} = \frac{N_{A}X_0\rho}{A} \int_0^T dt\,\int dE \, I(E,\,E_B,\,t)\,\sigma(E),
\ea
where $N_{e^+\mathrm{oT}}$ is the number of positron on target, $N_A$ is Avogadro's number, $X_0$ and $A$ are respectively the radiation length and the atomic mass of the target material,   and $\rho$ 
its  density. For the NA64 ECAL active target, neglecting the contribution 
of the plastic scintillator layers, we have 
$X_0=0.56$~cm, $\rho=11.35$ g/cm$^3$ and $A=207.2\,$.  We  
 focus on di-muon production in the 
pre-shower region, and take $T=4$.\footnote{We checked that after $T=4$ only a negligible amount of di-muons is produced.}
The function $I(E,\,E_B,\,t)$ gives the probability of finding a positron with energy $E$ after traversing $t$ radiation lengths~\cite{Bethe:1934za,Tsai:1966js} and is given by:
\begin{equation}
    I(E,\,E_B,\,t) = \frac{\theta(E-E_B)}{E\,\Gamma(bt)}\left(\log\frac{E}{E_B} \right)^{bt-1},
\end{equation}
where $b = 4/3$  and $\Gamma$ is the gamma function. 

It is worth noting that both the target’s thickness and, in particular, its high atomic number contribute to smearing the center-of-mass energy distribution -- the former through positron energy loss, and the latter due to the significant motion of inner-shell electrons. 
In comparison, the $\sim 1\%$ beam energy spread is entirely negligible and,
to expedite the integrations, it has  been omitted in the derivation of the full numerical results.
By contrast, in the FEAR approximation, the beam energy spread induces a noticeable 
smoothing of the differential event distribution and has therefore been explicitly included in that case. It is also worth noting that, given the positron beam intensity available at CERN NA 
($ N_{e^+\mathrm{oT}}\sim \mathcal{O}(10^{11}-10^{12})$) 
the maximum atomic electron momenta that can be probed remain below 
$\sim 10\,$MeV,  due to the strong suppression of the probability distribution at higher momentum values.
This corresponds to a length scale exceeding approximately $100\,$fm, which is safely larger 
than the radius of the lead nucleus ($\sim 6.6\,$fm). Therefore, nuclear effects 
can be neglected in the $e^+e^-$ annihilation process.

\section{Theoretical uncertainties}  
\label{sec:uncertainties}
In order to estimate the uncertainty in the predicted number of di-muon events from positron annihilation, we accounted for several effects. First, we computed $n(k_A)$ using different strategies: (i) by interpolating the CP at momenta below $\sim78$\,a.u. using spline functions; (ii) by fitting the data with polynomial curves, under the requirement that the resulting CP function be $\mathcal{C}^2$-continuous. These two methods provide a smooth derivative of the function $J(k_A)$. 
Second, we estimated $n(k_A)$ for $k_A>78$\,a.u. using results from the DBSR-HF code. The shape resulting from this calculation agrees with the shape of $n(k_A)$  obtained by fitting the experimental CP of Ref.~\cite{Pattison_1979} with either of the two methods (i) or (ii) outlined above.
However, the absolute values of the momentum distribution 
obtained from the DBSR-HF code overestimate the experimental values by approximately $50\%$.
%
%
We have thus estimated the theoretical uncertainty on the total number of events by comparing the values of $N_{\mu\mu}$ obtained using the DBSR-HF approach for $n(k_A)$ at $k_A > 78$ a.u., both directly and after rescaling to match the experimental data from  Ref.~\cite{Pattison_1979}.
Additionally, the total number of events was calculated using two independent Monte Carlo integrators: \texttt{SUAVE} from the \texttt{CUBA} package~\cite{Hahn:2004fe} implemented in \texttt{Mathematica}, and \texttt{VEGAS}~\cite{Lepage:1977sw,Lepage:2020tgj} for \texttt{python}. 
These comparisons suggest that the estimated uncertainty in the total number of predicted di-muon events does not exceed the 10\% level. 

\begin{figure*}[t!]
    \centering
        \includegraphics[width=0.75 \textwidth]{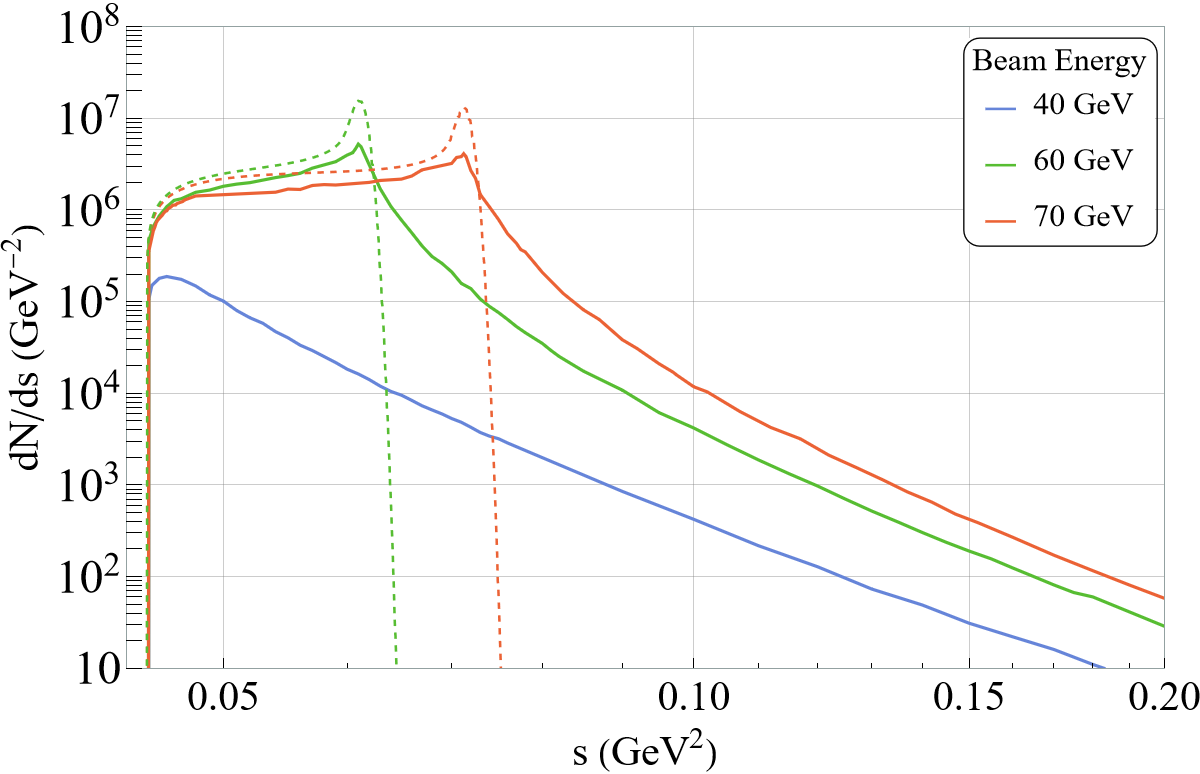}
    \caption{Number of $\mu^+\mu^-$ events produced at  NA64 
    via  $e^+e^-$ annihilation for positron beam energies of $E_B = 40$, 60, and 70\,GeV (shown as blue, green, and red solid lines, respectively) compared to the expected yields in the FEAR approximation (green and red dashed lines).  
    In this approximation,  for $E_B = 40\GeV$ the process remains below the c.m. energy threshold.
    }
    \label{fig:NA64}
\end{figure*}

\section{Results}  
\label{sec:results}
Figure~\ref{fig:NA64} shows the differential number of di-muon events as a function  of the c.m. energy square $s$. 
The blue curve corresponds to the differential number of  di-muon events produced via the annihilation channel for a beam energy $E_B=40$~GeV. 
This energy lies below the threshold of $E_{\rm th} \simeq 43.7$~GeV required for di-muon production via positron annihilation
off stationary electrons. Therefore, no events of this type are expected when adopting the FEAR approximation. 
However, as the blue curve clearly shows, in head-on collisions with positrons, atomic electrons can boost the c.m. energy above the threshold for di-muon production, resulting in a substantial event yield. The shape of the events distribution, which decays exponentially at large $s$, reflects the probability density of the atomic electron momentum. 

The green curve shows the differential event distribution for $E_B=60$~GeV, while the red curve corresponds to $E_B=70$~GeV. These distributions peak at $s\simeq 0.061$~GeV$^2$ and $0.072$~GeV$^2$, respectively. 
The peak values are consistent with  muons being predominantly produced in the initial  layers of the target and with the atomic electron momentum distribution peaking at small momenta,  which 
results in a larger probability of interaction with \emph{slow} electrons.
The portion of the curve to the left of each peak arises primarily from positron energy losses, and to a lesser extent from parallel-momentum electron–positron interactions that imply lower c.m.
energies.
%
Similar to the $ E_B =40\,$GeV case, the shape of the distribution to the right of the peaks in the green and red curves is essentially determined by the momentum density of atomic electrons in the target. At large values of $s$, the differential event distribution closely reflects the atomic electron momentum distribution, resulting in the blue, green, and red curves exhibiting approximately the same slope.
\begin{table}[ht]
    \centering
    \renewcommand{\arraystretch}{2.}
    \begin{tabular}{|c|c|c|}
    \hline
     $E_B/$GeV  & FEAR  & AEM \\
    \hline
      \rule[-10pt]{0pt}{5pt} 40   & $0\,\,(E_B<E_\mathrm{th})$  & $1.4\times 10^3 \left(\dfrac{N_{e^+oT}}{10^{11}}\right)$ \\
    \hline
      \rule[-10pt]{0pt}{5pt} 60   & $7.2\times 10^4 \left(\dfrac{N_{e^+oT}}{10^{11}}\right)$  & $5.0\times 10^4 \left(\dfrac{N_{e^+oT}}{10^{11}}\right)$ \\   
    \hline
      \rule[-10pt]{0pt}{5pt} 70    & $\:\:9.2\times 10^4 \left(\dfrac{N_{e^+oT}}{10^{11}}\right)\:\:$  & $\:\:6.4\times 10^4 \left(\dfrac{N_{e^+oT}}{10^{11}}\right)\:\:$ \\
    \hline
    \end{tabular}
    \caption{Predicted number of $\mu^+\mu^-$ events in the FEAR approximation  and including the atomic electron motion (AEM) for $E_B=40,\,60$  and $70$ GeV.}
    \label{tab:summary}
\end{table}

The total number of events obtained by integrating the three distributions is summarized in Table~\ref{tab:summary}. In what follows, we adopt a benchmark number of positrons on target of $N_{e^+oT}=10^{11}$, corresponding to two-week NA64 runs in 2028 (at $60$ GeV) and 2029 (at $40$ GeV). Future data-taking in 2030-2031 is expected to increase this dataset by a factor of five~\cite{Bisio:2887649}. The results for different values of $N_{e^+oT}$  can be obtained by simple rescaling, as indicated in Table~\ref{tab:summary}. 

With $E_B=40$ GeV, NA64 is expected to detect around $1.4\times 10^3$ di-muon events from $e^+ e^- $ annihilation, while no signal above background is anticipated under the stationary electron approximation, as the beam energy lies below the di-muon production  threshold via the annihilation channel.  Increasing the beam energy to $60$ GeV ($70$ GeV),  the FEAR approximation  predicts around $7.2\times 10^4$ ( $9.2\times 10^4$) events. However, when electron motion effects are properly accounted for,   the expected number of events drops to about 
 $ 5.0\times 10^4$ ($6.4\times 10^4$),  corresponding approximately to 
 a 30\% reduction. 
The reason for this reduction can be easily  identified by examining the blue curve ($E_B = 40\,$GeV) in Fig.~~\ref{fig:NA64}. 
It shows that, for $s\lesssim 0.05$, di-muon production is not significantly suppressed.
If electrons were assumed to be at rest, this c.m. energy would correspond to a beam energy of approximately  $E_B \simeq 50\,$GeV. 
We can thus conclude that the  region in the electron momentum distribution in which the probability remains of the order of its zero-momentum value corresponds, within the FEAR approximation, to an effective beam energy spread of the order of $|\Delta E_B| \approx 10\,$GeV.
From Fig.~\ref{fig:FEARxs}, we see that within an interval of this size, the di-muon production cross section in the FEAR approximation is close to its maximum for both $E_B = 60\,$GeV and $E_B = 70\,$GeV. 
Therefore, the smearing of the c.m. energy induced by the motion of target electrons, which can be qualitatively simulated 
by assuming for the cross section in Fig.~\ref{fig:FEARxs}
an effective beam energy broadening of the order of 10\,GeV, can only act to reduce the total yield of di-muon events.
Thus, for beam energies above threshold, and not far beyond the maximum of the production cross section,  
a sizable reduction in the number of di-muon events is predicted to be a generic phenomenon. Although less striking than di-muon production in $e^+e^-$ annihilation with a beam of seemingly sub-threshold energy, the observation of such reductions would provide an equally compelling signature of electron motion effects in high-energy collisions.

\bigskip

\section{Conclusions}
\label{sec:conclusions}
The effects of atomic electron motion within materials provide a powerful means for extracting information about atomic electron wavefunctions, and are therefore of fundamental importance in chemistry and condensed matter physics. For several decades, these effects have been extensively studied using low-energy probes,  for example in Compton scattering with X-rays~\cite{Williams:01011983} and in positron annihilation spectroscopy~\cite{Ghosh:2000}.
Conversely, in the analysis of high-energy fixed-target experiments,
it is generally  implicitly assumed that atomic electrons can 
be treated as free and at rest when participating  
in  scattering interactions with  an incident high-energy beam.
As a result, to date, these effects have been systematically neglected in 
most cases.\footnote{It is worth noting that an approximate treatment of electron motion effects in positron annihilation on target electrons has recently been incorporated into a development~\cite{Oberhauser:2024ozf} of the GEANT4-compatible package DMG4~\cite{Bondi:2021nfp}, used for simulating light dark matter production in fixed-target experiments.}
Given the theoretical accuracy required by current and upcoming precision experiments involving neutrino, electron, positron, and muon scattering off atomic electrons in fixed targets, it has now become essential to incorporate corrections beyond the FEAR approximation.
These include, most notably, the effects of atomic binding energies~\cite{Plestid:2024xzh}; electromagnetic corrections, which, particularly in high-$Z$ materials, can be significantly enhanced by the large electric charge of the atomic target and final-state debris~\cite{Plestid:2025ojt,Plestid:2024jqm}; and atomic electron motion effects~\cite{Arias-Aragon:2024qji,Arias-Aragon:2024gpm,Arias-Aragon:2025kiz}.
Recent studies have started addressing these issues in detail for specific processes. 
In particular, the inclusion of electron motion effects in the search  
for  hypothetical massive $X$ bosons 
via the resonant process  $e^+e^- \to X \to e^+e^-$  
has been shown to be essential for accurately modeling the signal 
shape~\cite{Arias-Aragon:2024qji}.  In high-$Z$ materials, 
electrons in the inner shells are characterized by particularly  
high momenta, and this feature can be leveraged to 
search for new particles via the resonant $e^+e^-$ annihilation process,  
extending the accessible mass range well beyond what the naive FEAR approximation would suggest~\cite{Arias-Aragon:2025kiz}. Moreover, with the aid of high-intensity positron beams, such as the 12\,GeV beam under development at the Continuous Electron Beam Accelerator Facility (CEBAF) 
at JLab~\cite{Afanasev:2019xmr,Accardi:2020swt,Arrington:2021alx}, it may become possible to probe the tail of the electron momentum distribution and measure the energy dependence of the hadronic cross section in $e^+e^-$ annihilation up to c.m. energies around 1\,GeV~\cite{Arias-Aragon:2024gpm}, thereby offering an alternative data driven approach to determine the hadronic vacuum polarization contribution to the muon $g-2$.

Despite  these promising opportunities,  the study of atomic electron momentum effects in high-energy collisions has, to date, remained mostly theoretical, as no clear experimental evidence has yet been reported in such settings. In this work, we have shown that the NA64 experiment at CERN offers a unique opportunity to directly observe the effects of atomic electron motion in high-energy positron annihilation off target electrons.
By incorporating the atomic electron momentum distribution into the calculation of 
di-muon production from $e^+e^-$ annihilation, we have shown that a $40$\,GeV positron beam  --- nominally below the di-muon production threshold  for stationary electrons,  $E_{\rm th} \simeq 43.7$\,GeV --- 
is nevertheless expected to yield a sizable number of di-muon events. 
The detection of these events at NA64 would constitute a direct, striking evidence of bound-electron momentum contributions to the c.m. energy of the collision. 
We have also explored the effects of atomic electron motion on di-muon production for positron beam energies well above threshold ($60$\,GeV and $70$\,GeV). We have shown that properly accounting for electron motion leads to a sizable broadening of the collision's c.m. energy distribution,  reducing the predicted di-muon yield  by approximately  $30\%$  relative to the FEAR approximation. 

In conclusion, our results highlight NA64’s potential to provide a first direct high-energy confirmation of atomic electron momentum effects. Such a measurement would strengthen 
the physics case for leveraging electron motion in  high--$Z$ materials to  
raise the collision c.m. energy, effectively exploiting the initial states of beam positrons and atomic electrons in a manner somewhat analogous to an asymmetric collider.
Future dedicated analyses of NA64 data could thus open a novel pathway to 
exploring the interplay between low-energy studies of electronic  structures of materials and high-energy particle physics.

\acknowledgments
We thank  
Andrea Celentano,  Mikhail Kirsanov and Luca Marsicano
for several discussions.
We warmly thank Lau Gatignon and Johannes Bernhard for providing detailed information on the CERN NA beams, and Eric Voutier for drawing our attention to the SLAC measurement of the Levchuk effect. 
We gratefully acknowledge Luc Darmé for his valuable contributions to the study of atomic electron momentum effects.
F.A.A., G.G.d.C. and E.N. are supported in part by the INFN ``Iniziativa Specifica" Theoretical Astroparticle Physics (TAsP). F.A.A. received additional support from an INFN Cabibbo Fellowship, call 2022.
G.G.d.C. acknowledges LNF for hospitality at various stages of this work. The work of E.N. is also supported  by the Estonian Research Council grant PRG1884.   Partial support from the CoE grant TK202 “Foundations of the Universe” and from the CERN and ESA Science Consortium of Estonia, grants RVTT3 and RVTT7, and from the COST (European Cooperation in Science and Technology) Action COSMIC WISPers CA21106 are also acknowledged.

\bibliographystyle{JHEP}
\bibliography{biblio.bib}

\end{document}